\documentstyle[pre,aps,twocolumn,floats]{revtex}

\begin{document}
\title{Dielectric response of cylindrical nanostructures in a\\ magnetic field}

\author{Peter Fulde and Alexander Ovchinnikov}  

\address{
Max-Planck-Institut f\"ur Physik komplexer Systeme,
N\"othnitzer Stra\ss e 38,
01187 Dresden (Germany)}

\date{\today}
\maketitle

\abstract

We study the magnetic field dependence of the dielectric response of large
cylindrical molecules such as nanotubes. When a field-induced level crossing
takes place, an applied electric field causes a linear instead of the usual
quadratic Stark effect. This results in a large dielectric response. Explicit
calculations are performed for doped nanotubes and a rich structure in the real
part of the low-frequency dielectric function $\epsilon'(H)$ is found when a
magnetic field is applied along the cylinder axis. It is suggested that studies
of $\epsilon'(H, T)$ can serve as a spectroscopic tool for the investigation of
large ring-shaped or cylindrical molecules.\\[8ex]

\noindent PACS.~~~~~~~~\begin{minipage}[t]{15cm}
77.22.-d Dielectric properties of solids and liquids\\
78.40.Ri Fullerenes and related materials\\
75.20.-g Diamagnetism and paramagnetism\\
73.23.-b Mesoscopic systems
\end{minipage}

\vspace{1.5cm}

Dedicated to K. Biedenkopf on the occasion of his 70th birthday.

\newpage

\newcommand{\gsim}{\mathrel{\raise.3ex\hbox{$>$\kern-.75em\lower1ex\hbox{$\sim$}}}}
\newcommand{\lsim}{\mathrel{\raise.3ex\hbox{$<$\kern-.75em\lower1ex\hbox{$\sim$}}}}

{\section{\bf Introduction}}

During the last years considerable progress has been made in precise
measurements of the real part of the low-frequency dielectric function
$\epsilon'(\omega)$. In particular those measurements could be extended to
ultra-low temperatures, i.e., down to a few mK. Ratios of $\delta
\epsilon'/\epsilon'$ up to $10^{-7}$ were achieved in that temperature
regime. A rather spectacular success associated with that progress was the
observation of a strong magnetic-field dependence of the polarizability of
multicomponent glasses in the mK regime \cite{ref1,ref2}. This development
suggests reconsideration of magnetic field effects on ring molecules or related
structures like nanotubes. In both cases an applied magnetic field induces a
diamagnetic ring current. Due to this current the energy of the ground state
increases quadratically with the applied magnetic field. This continues until
one of the excited states which lowers its energy in  a field crosses the
ground-state and becomes the new ground state. At the crossing point an applied
electric field causes a linear Stark effect, instead of the usual quadratic one
and hence a divergent electric polarizability. The physical origin of the
crossover is easily understood. An excited state carrying a ring current in the
absence of a magnetic field becomes a state without a ring current, when a
sufficiently high magnetic field is applied, because the induced current may
cancel the original one. When this is the case the energy of that state equals
the one of the ground state in the absence of a field. This simple argument
shows that the ground-state energy is a periodic function of an applied
magnetic field. The periodicity is given by the flux enclosed by the ring
current. When one flux quantum $\phi_0 = hc/e$ is penetrating the ring the
ground-state energy has returned to its original value. It is well known that a
huge field of order $10 ^5~ T$ is needed for a flux quantum $\phi_0$ to
penetrate a benzene molecule consisting of a ring of six carbon atoms. Because
of that large value of the magnetic field, effects of it on ring molecules have
obtained only little attention in the past \cite{ref3,ref4}. The purpose of the
present paper is to point out that the situation has changed considerably. Not
only has there been experimental progress in performing high precision
measurements of $\epsilon'$ at low temperatures, but also the synthesis of
organic ring structures has made impressive advances. For example, nanotubes of
large circumference have been produced, lowering the field required for the
enclosure of a flux unit. In this paper we want to demonstrate that a
measurement of $\epsilon'(H, T)$ should provide important information on the
electronic excitations of ring molecules, in particular on level crossings. As
a first step, we calculate here the dielectric response of a molecule
consisting of a square lattice, e.g., of carbon sites bent into a cylindrical
form and of a nanotube, i.e., a bent honeycomb lattice. Various extensions of
the work presented here will follow later. 

The magnetic and electric field are assumed
to be directed along the cylindrical axis. The electron interactions are
assumed to be included in effective one-electron parameters like in an extended
Hückel theory or in the quasiparticle theory of Landau. In a subsequent
investigation we shall include the electronic interactions more explicitely
than done here. This may have profound effects on the results. A rich structure
in $\epsilon'(H, T)$ is obtained which should be experimentally observable. It
is closely related to the low-energy excitations of the systems in an applied
magnetic field and in particular to level crossings as the field
changes. Although our findings are limited here to the cylindrical structures
described above, they suggest detailed experimental studies of $\epsilon'(H,
T)$ for ring molecules. This seems to us a largely untouched field of research.

In order to explain the main features of $\epsilon'(H, T)$ we consider first a
single ring of $N$ sites in a magnetic field along the ring axis. The
Hamiltonian is

\begin{equation}
\label{1} H = t \sum_{n, \sigma} (a^+_{(n+1)\sigma}~ a_{n \sigma}~ e^\frac{2
\pi i \phi}{N} + {\rm h.c.})
\end{equation}

\noindent where $a^+_{n \sigma}, a_{n \sigma}$ are electron creation and
annihilation operators and $\phi$ is the magnetic flux through the ring in
units of the flux quantum $\phi_0$. The resulting energy eigenvalues are

\begin{equation}
\label{2} \epsilon (q) = 2t~ {\rm cos} \left[ \frac{2 \pi}{N} (q +
\phi) \right]~~,~~~~~~~~~~~q = 0, \pm 1, \pm 2, ... 
\end{equation}

The ground-state energy $E_g$ is periodic in the flux, i.e., $E_g (\phi + 1) =
E_g (\phi)$. More explicitely we write for $E_g$

\begin{eqnarray}
\label{3} E_g (\phi) & = & \sum_{\rm occ} 2t~ \left[ {\rm cos} \left( \frac{2
\pi}{N} q \right)~ {\rm cos} \left( \frac{2 \pi}{N} \phi
\right)\right.\nonumber \\ 
&&\left. - {\rm sin}
\left( \frac{2 \pi}{N} q \right)~ {\rm sin} \left( \frac{2 \pi}{N} \phi
\right)\right]~~. 
\end{eqnarray}
 
For a closed-shell system, i.e., for an electron number $N_e = 4n + 2$, where
$n$ is an integer we find that

\begin{equation}
\label{4} \sum_{\rm occ} {\rm sin} \frac{2 \pi}{N} q = 0~~.
\end{equation}
 
\noindent In that case the field-dependent contribution to the ground-state
energy is 

\begin{eqnarray}
\label{5} \delta E_g (\phi) & = & \sum_{\rm occ} 2t~ {\rm cos} \left( \frac{2
\pi}{N} q \right) \left( {\rm cos} \left( \frac{2 \pi}{N} \phi \right) -1
\right)\nonumber \\ 
& = & -E_g (0) \left( 1 - {\rm cos} \left( \frac{2 \pi}{N} \phi \right)
\right)~~. 
\end{eqnarray}
 
For large $N$ we may expand this expression and obtain

\begin{equation}
\label{6} \delta E_g (\phi) = -E_g (0)~ \frac{2 \pi^2}{N^2}~ \phi^2 > 0
\end{equation}
 
\noindent for $\phi \leq \frac{1}{2}$. When $\phi = \frac{1}{2}$ the ground
state is twofold degenerate because of a level crossing at that point. For
$\phi > \frac{1}{2}$, the expression (3) is replaced by

\begin{equation}
\label{7} \delta E_g (\phi) = -E_g (0)~ \frac{2 \pi^2}{N^2}~ (1 -
\phi)^2~~,~~~~~~~~~\phi > \frac{1}{2}~~.
\end{equation}
 
The behaviour of $\delta E_g (\phi)$ is schematically shown in Fig. 1. The
contribution of $\delta E_g (\phi)$ to $E_g (\phi)$ is very small for large
values of $N$ and hardly detectable. This does not hold true though for other
quantities. For example, when in addition an electric field is applied
perpendicularly to the ring axis its effect on the ground-state energy is
strongly dependent on $\phi$. For $\phi = \frac{1}{2}$ we are dealing with a
linear Stark effect instead off a quadratic one when $\phi \neq
\frac{1}{2}$. Therefore the dielectric constant has a singularity at that
particular value of $\phi$. This simple example sheds light onto the physical
reason why the dielectric function can be so sensitive to an applied magnetic
field. The same feature is found for cylindrical molecules which are subject of
this paper. 


\vspace{1cm}

{\section{\bf Magnetic field dependence of the free energy}}

In order to demonstrate the influence of an applied magnetic field $\bf{H}$
on the free energy we consider two different systems. One is a model square
lattice rolled into the form of a cylinder. The other one is a nanotube which
consists of a honeycomb lattice rolled into a cylinder in the same way.

We start with the square lattice forming a cylinder. It consists of $N$ atoms
along the perimeter and of $N$ atoms along the cylindrical axis z. The
eigenvalues depend on the flux $\phi$ though the cylinder and are of the form

\begin{eqnarray}
\label{8} \epsilon_{pq\sigma} & = & -{\rm cos} \left[ \frac{2 \pi}{N}~ (p +
\phi)\right] -t{\rm cos} \left[ \frac{2 \pi}{M + 1}~q \right]\nonumber \\
&& + 2 \pi^2~
\frac{m_{\rm eff}\sigma}{m}~ \frac{\phi}{N^2}~~.  
\end{eqnarray} 

The first term corresponds to a transfer integral of size $-\frac{1}{2}$ along
the perimeter and the second to one of magnitude $-\frac{t}{2}$ along the z
axis. The parameters $p$ and $q$ take the integer values $p = 1, ..., N$ and
$q = 1, ..., M$, respectively. The last term is the Zeeman contribution which
is expressed here in terms of the flux $\phi$. Since the latter is in units
of the flux quantum $\phi_0$, the ratio of the effective mass
$m_{\rm eff}$ divided by the electron mass $m$ enters here, with the former
referring to an electronic motion perpendicular to the z axis. Furthermore,
$\sigma = \pm 1$. The free energy of the system is of the usual form

\begin{equation}
\label{9} \beta F = - \sum_{p q \sigma}~ {\rm ln} [1 + {\rm exp} (-\beta
(\epsilon_{pq\sigma} - \mu))] 
\end{equation}

\noindent where $\beta = (k_B T)^{-1}$ and $\mu$ is the chemical potential. It
is determined by expressing the number of electrons $N_e$ in terms of it, i.e.,

\begin{equation}
\label{10} N_e = \sum_{pq\sigma} \frac{1}{{\rm exp}[\beta(\epsilon_{pq\sigma} -
\mu)]+ 1}~~. 
\end{equation}

In practice we calculate $\mu$ by first choosing an approximate value $\mu_0$
and calculating the corresponding value $N_e^{(1)}$. The correction
$\delta\mu_0$ to $\mu_0$ can then be determined from

\begin{equation}
\label{11} \delta \mu_0 = -\frac{1}{\beta} {\rm ln} \left( 1 + \frac{N_e^{(1)} -
N_e}{a(T)}\right)~~. 
\end{equation}

This expression is more convenient for numerical calculations than its
linearized version in $(N_e^{(1)} - N_e)/a(T)$ where the function $a(T)$ is
given by 

\begin{equation}
\label{12} a(T) = \frac{1}{4} \sum_{pq\sigma} {\rm cos}h^{-2}~ \left[\frac{\beta}{2}
(\epsilon_{pq\sigma} - \mu_0)\right]~~. 
\end{equation}

One can use the corrected potential $\mu_1 = \mu_0 + \delta\mu_0$ in order to
calculate the next correction $\delta\mu_1$. We obtain with $\delta\mu_1$ the
chemical potential already with an accuracy of order $N^{-2} M^{-2}$, which
perfectly serves our purposes. 

The same procedure can be applied to carbon nanotubes. In that case the unit
cell contains four carbon atoms. Hence the excitation energies form four bands,
i.e.,

\begin{eqnarray}
\label{13} \epsilon_{pq\sigma} (\phi) & = & \pm \left( 1 + u_p \pm (1 + u_p
v_q)^\frac{1}{2} \right)^\frac{1}{2}\nonumber \\
&& +~ 2 \pi^2 \frac{m_{\rm eff}}{m}~ \frac{\sigma\phi}{N^2} 
\end{eqnarray}

\noindent with

\begin{eqnarray}
\label{14} u_p & = & 2 \left(1 + {\rm cos} \left[ \frac{2\pi}{N} (p +
\phi)\right] \right)\nonumber \\ 
v_q & = & 2 \left(1 + {\rm cos} \left[ \frac{2\pi}{M+1} q\right] \right) 
\end{eqnarray}

\noindent and $p = 1, ....., N;~~q = 1, ....., M$ \cite{ref5,ref6}.

\vspace{1cm} 

{\section{\bf Induced dipole moment}}

When an electric field is applied along the $z$ axis the excitation spectrum of
the system can no longer be calculated exactly. Instead, approximations have to
be made. We cannot apply linear response theory because of a linear Stark
effect at level crossings. Since in practise the applied electric field is very
small, the density of electrons changes only slightly along the z axis. This
enables us to determine the induced density changes by using a quasiclassical
approximation. Within that scheme the excitation energies depend not only on
$p, q$ and $\sigma$ but on the coordinate z as well. We illustrate the
approximation by considering a chain of $M$ atoms as a simple example. The
Hamiltonians is of the form

\begin{eqnarray}
\label{15} H_{\rm 1d} & = & - \sum^{M-1}_{n=1, \sigma} (a^+_{(n+1)\sigma}
a_{n\sigma} + {\rm h.c.})\nonumber \\
&& +~ e a_0 F_0~\sum_{n\sigma} a^+_{n\sigma} a_{n\sigma} \left( n-\frac{M+1}{2}
\right)~~. 
\end{eqnarray} 

Here $a_0$ is the lattice constant and $F_0$ is an applied electric field along
the chain direction.

Exact calculations of the induced dipole moment D require an evaluation of
the expression

\begin{eqnarray}
\label{16} {\rm D} & = & \frac{Sp~ \hat{d}~ e^{-\beta(H_{\rm 1d} - \mu)}}{Sp~
e^{-\beta(H_{\rm 1d} - \mu)}}\nonumber \\
& = & \sum^M_{k=1,\sigma} \frac{(\hat{d})_{kk}}{1 + e^{\beta(E_{k\sigma} -
\mu)}}~~. 
\end{eqnarray}

\noindent Here $\hat{d}$ is dipole operator

\begin{equation}
\label{17} \hat{d} = e a_0 \sum^M_{n=1,\sigma} a^+_{n\sigma} a_{n\sigma}
\left( n - \frac{M+1}{2} \right) 
\end{equation} 

\noindent and $E_{k\sigma}$ denotes the excitation energies of the chain. In
order to compute D from (\ref{16}) we have to diagonalize $H_{\rm 1d}$ in
order to find the eigenenergies and eigenfunctions of that Hamiltonian. This
can be done if not more than 1000 atoms are involved. Instead of doing that
we want to use here a simpler, more effective quasiclassical scheme. In the
quasiclassical approximation the excitation spectrum is of the form

\begin{equation}
\label{18} \epsilon_{pm} = - 2 {\rm cos} \left[ \frac{2\pi}{M+1}~ p\right] + e
a_0 F_0 \left( m - \frac{M+1}{2} \right) 
\end{equation} 

\noindent with $p = 1, ....., M$ and $m = 1, ....., M$. The corresponding
expression for the induced dipole moment is

\begin{equation}
\label{19} {\rm D} = \frac{2e a_0}{M} \sum_{pm} \frac{(m -
(M+1)/2)}{e^{\beta(\epsilon_{pm} - \mu)} + 1}~~. 
\end{equation} 

We have calculated D for a chain of $M = 201$ atoms by using (\ref{16}) and
alternatively (\ref{18}). The results are compared in Fig. 2 for different
densities and temperatures. The deviations caused by the semiclassical
approximation  are less than $1 \%$ or $\frac{1}{M}$ in all cases. This
justifies the use of a quasiclassical approximation when we calculate the
dielectric response of cylindrical molecules such as nanotubes in an applied
magnetic field. 

\vspace{1cm} 

{\section{\bf Results and discussions}}

In the following we want to present results for the dielectric response of the
two types of cylindrical molecules described above, i.e., for a square lattice
rolled into a cylinder and for nanotubes. The induced dipole moment is
calculated in close analogy to the one of a ring, although here the electric
field $F_0$ is directed along the cylindrical axis. We start with the
square-lattice case. In analogy to (\ref{19}) the induced dipole is calculated
from

\begin{equation}
\label{20} {\rm D} (\phi) = \frac{e~ a_0}{M} \sum_{mpq\sigma}~ \frac{(m -
(M+1)/2)}{e^{\beta [\tilde{\epsilon}_{pq\sigma} (m, \phi) - \mu(\phi)]} + 1}
\end{equation} 

\noindent where

\begin{equation}
\label{21} \tilde{\epsilon}_{pq\sigma} (m, \phi) = \epsilon_{pq\sigma} (\phi) +
e a_0 F_0~ (m - (M+1)/2) 
\end{equation} 

and $\epsilon_{pq\sigma}(\phi)$ is given by (\ref{8}). Results for the
magnetic-field dependent part ${\rm D}(\phi) - {\rm D}(0)$ are shown in Fig. 3
for a cylinder with 100 atoms along the circumference and 1000 atoms along the
axis, i.e., $N = 100$ and $M = 1000$, respectively.  Note that $({\rm D}(\phi)
- {\rm D}(0))/{\rm D}(0) = (\epsilon'(\phi) - \epsilon'(0))/ \epsilon'(0)$
where $\epsilon'$ is the real part of the dielectric response in the
low-frequency limit. The temperature, or more precisely $k_B T$ is $10^{-4}$ in
units of the hopping matrix element. The chosen density corresponds to 0,74
electrons per site. One notices a rich structure as a function of the applied
magnetic field. 

A cylinder formed from a square lattice is a hypothetical case. But if one
assumes a lattice constant $a_0 = 1.40 {\AA}$ as in the case of an aromatic
carbon ring, the field required for the enclosure of a flux unit is of order
$260 T$. It is derived from the following relation between the flux $\phi$ (in
units of $\phi_0$) and the applied magnetic field

\begin{equation}
\label{22} \phi = \frac{N^2 a^2_0~ eH}{8\pi^2 \hbar c}~~. 
\end{equation} 

The structure in $\epsilon'(H)$ obtained within that range of fields reflects
properties of excited states, in particular crossings of energy levels.

For nanotubes the calculations are done quite similarly, but here we have to
take a sum over all four energy bands. The computational results are shown in
Fig. 4 for a density of $n = 0.89$ $\pi$-electrons per site. One notices that
the rich structure in $\delta\epsilon'(H)/\epsilon'(H)$ in the regime $0<H<40T$
is of order unity and therefore should be easily detectable. Results for
other densities look similar, except for $n = 1$ which is special. The reason
is that a honeycomb or graphite lattice has for $n = 1$ a Fermi surface
consisting of a point. Therefore in a a finite system the level spacing close
to the Fermi energy is particularly large. This leads to small changes in
$\delta\epsilon'(H)/\epsilon'(0)$ only. The situation changes at high magnetic
fields. Due to the Zeeman term in the Hamiltonian the spin dependent densities
$n_\sigma$ differ more and more from each other, i.e., $n = n_\uparrow+
n_\downarrow$ with $n_\uparrow (H) \neq  n_\downarrow (H)$ and the Fermi
surface moves away from the special point at half filling. This brings us back
to the doped case and we obtain again a rich structure in
$\delta\epsilon'(H)/\epsilon'(0)$ like in Fig. 4. 

\vspace{1cm}

{\section{\bf Conclusions}}

The above calculations show that large molecules of cylindrical or circular
shape should show detectable magnetic field effects due to the Bohm-Aharonov
effect. They lead to a strong variation of the dielectric function in the
low-frequency limit as function of the applied magnetic field. Those variations
are predominantly caused by doubly degenerate ground states resulting from
level crossings in the applied field. At a crossing point an applied electric
field causes a linear Stark effect instead of a quadratic one when the ground
state is nondegenerate. The present investigation requires a number of
extensions which will be the subject of separate investigations. One concerns
the dependence of $\delta\epsilon'(H)$ on the directions of the applied
magnetic and electric fields. Important is also a proper inclusion of electron
correlations. As pointed out before, the present calculations have been done
within the one-electron approximation.  But correlations, in particular when
they are strong will clearly result in important modifications of the
dielectric response \cite{ref7,ref8}. Finally, we also have to generalize the
above theory to the case of mutually interacting molecules. This may become an
important issue when platelike molecules are forming stacks and a magnetic
field is applied along the direction of the stack. Although the work presented
here needs extensions of the form just described it is fair to state that
the results presented here justify efforts towards a systematic investigation
of the magnetic-field dependent dielectric response of ring- or cylinder
shaped molecules. We feel that in the future they may develop into a
spectroscopic tool for studying low-energy excitations of such systems.

\newpage

\noindent \hspace{4cm} FIGURE CAPTIONS\\[-2ex]

\begin{footnotesize}
\newcounter{fig}
\begin{list}{Fig. \arabic{fig}:}{\usecounter{fig}
   \setlength{\labelwidth}{1.6cm} \setlength{\leftmargin}{1.8cm}
   \setlength{\labelsep}{0.4cm}   \setlength{\rightmargin}{0cm}
   \setlength{\parsep}{0.5ex plus0.2ex minus0.1ex}
   \setlength{\itemsep}{0ex plus0.2ex}}
\item Schematic plot of $\delta E_g(\phi)$ {\it vs.} $\phi$ (thick solid
line). The level crossings are due to parabolas describing different excited
states $\epsilon_i(\phi)$ shifted by flux units. When levels cross, an applied
electric field causes a linear Stark effect and hence induces a divergent
dielectric response. 
\item Induced dipole moment as a function of electrons per site for a chain of
$M = 201$ atoms calculated with the exact quantum-mechanical expression
(\ref{16}) (dashed lines) and when a semiclassical approximation (\ref{19})
(solid lines) is made.  (a) and (b) correspond to temperatures $k_B T = 0.01$
and $0.05$, respectively (in units of the transfer integral). 
\item Dielectric response $[\epsilon'(H) - \epsilon'(0)]/\epsilon'(0)$ for a
model square-lattice system with $N=200, M=1000$ in an axial magnetic
field. The temperature is $k_BT = 10^{-4}$ (in units of the transfer integral),
and $a_0 = 1.4 {\AA}$. The density is 0.74 electrons per site.
\item Dielectric response $[\epsilon'(H) - \epsilon'(0)]/ \epsilon'(0)$ for a
nanotube with $N=100, M=1000$ in an axial magnetic field. The temperature is
$k_B T= 10^{-4}$ in units of 3 eV and the density is 0.89 electrons per site.
\end{list}
\end{footnotesize}

\begin{thebibliography}{99}
\vspace{0.5cm}
\bibitem{ref1} P. Strehlow, C. Enss, and S. Hunklinger, Phys. Rev. Lett. {\bf
80}, 5361 (1998)
\bibitem{ref2} P. Strehlow, M. Wohlfahrt, A. G. M. Jansen, R. Haueisen,
G. Weiss, C. Enss, and S. Hunklinger, preprint 
\bibitem{ref3} {\it Electronic Structure Calculations on Fullerenes and their
Derivatives}, ed. by J. Cioslowski (Oxford University Press, New York, Oxford
1995), p. 139 
\bibitem{ref4} N. Hamada, S. Sawada, and A. Oshiyama, Phys. Rev. Lett. {\bf
68}, 1579 (1992) 
\bibitem{ref5} K. Tanaka, K. Okahara, M. Okada, and T. Yamabe,
Chem. Phys. Lett. {\bf 191}, 469 (1992)
\bibitem{ref6} A. A. Ovchinnikov, Phys. Lett. {\bf A 195}, 95 (1994)
\bibitem{ref7} F. V. Kusmartsev, Phys. Lett. {\bf A 161}, 433 (1992)
\bibitem{ref8} B. Sutherland, and B. S. Shastry, Phys. Rev. Lett.  {\bf 65},
1833 (1990) 
\end{thebibliography}
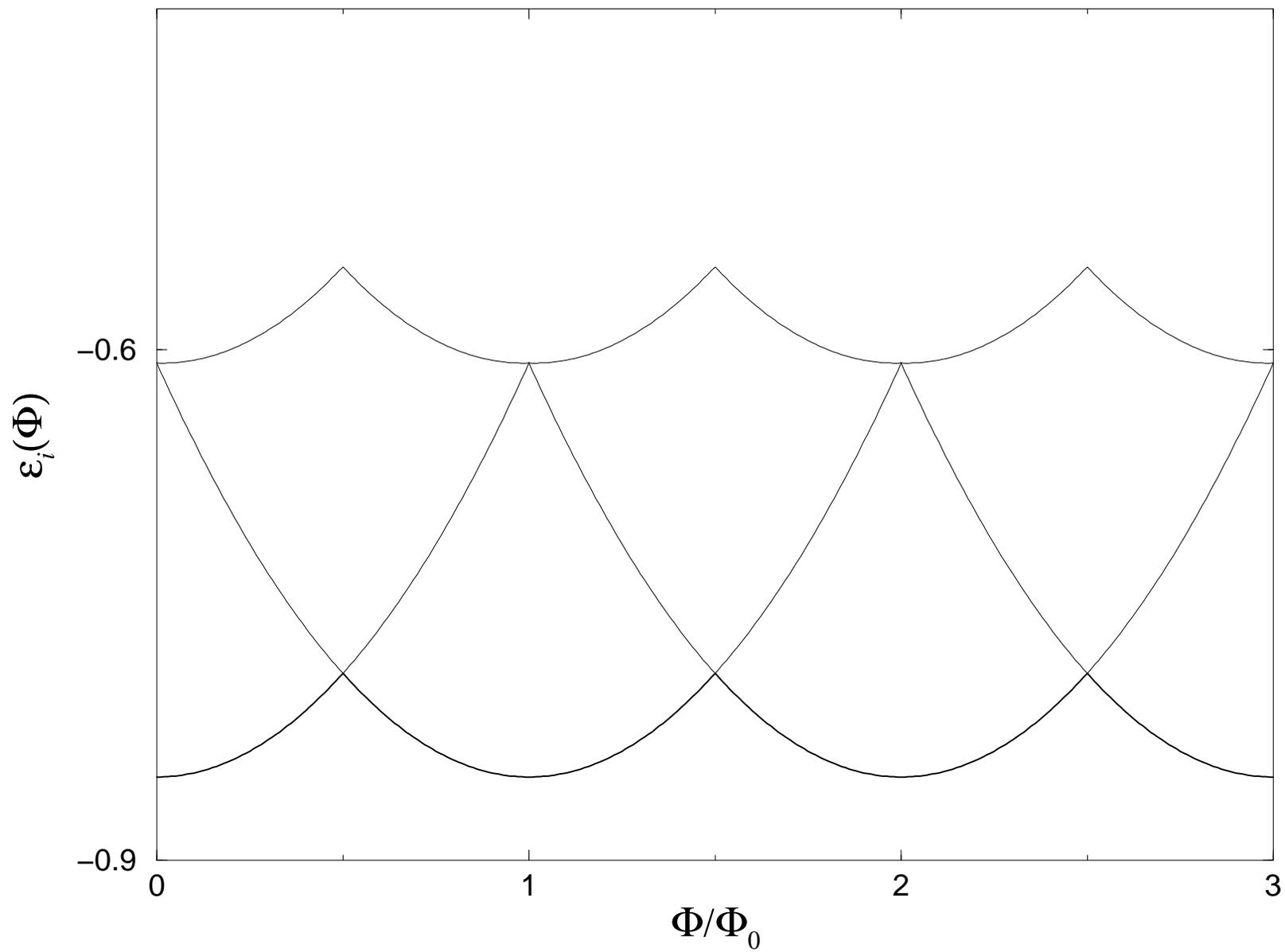
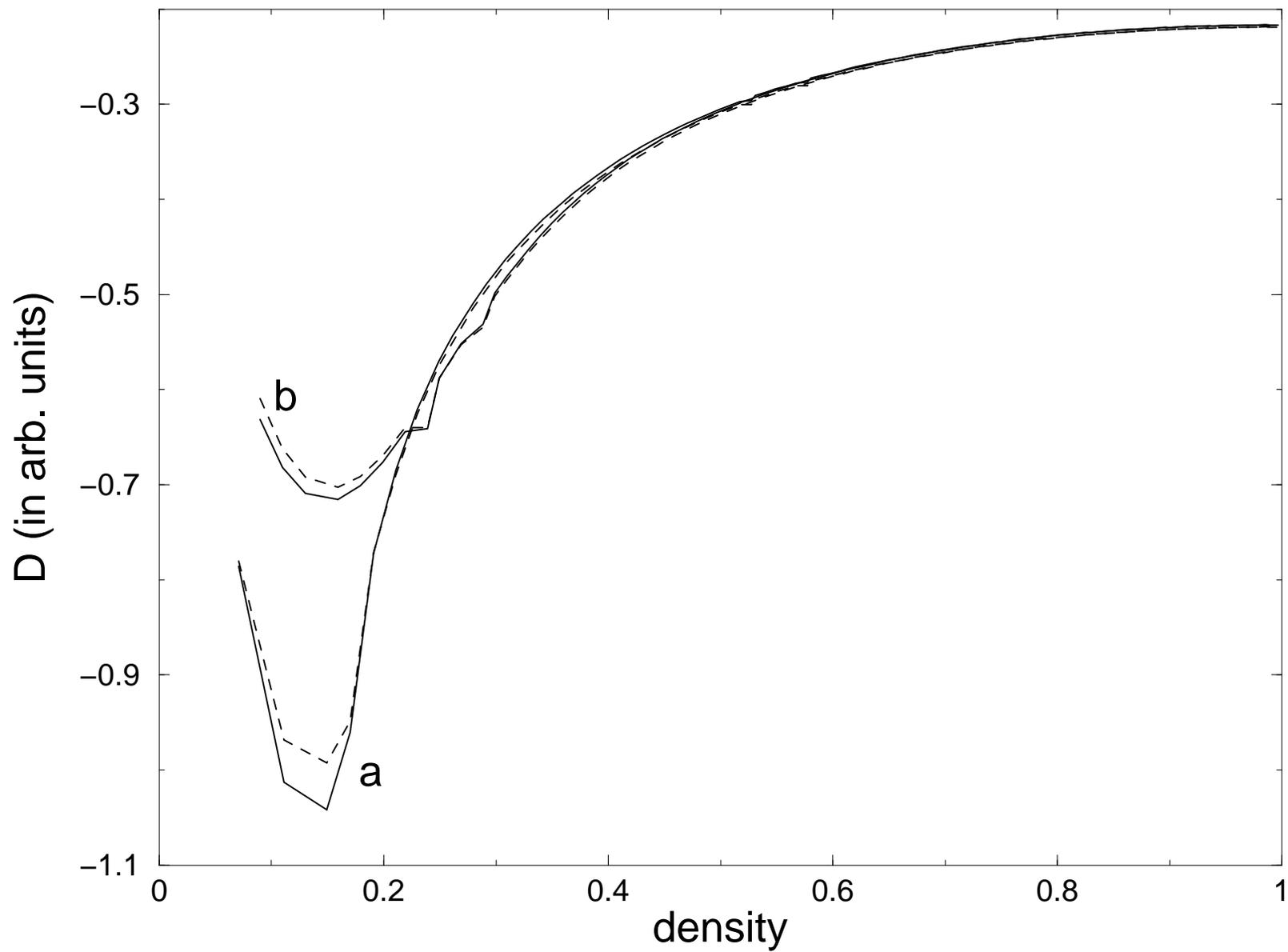
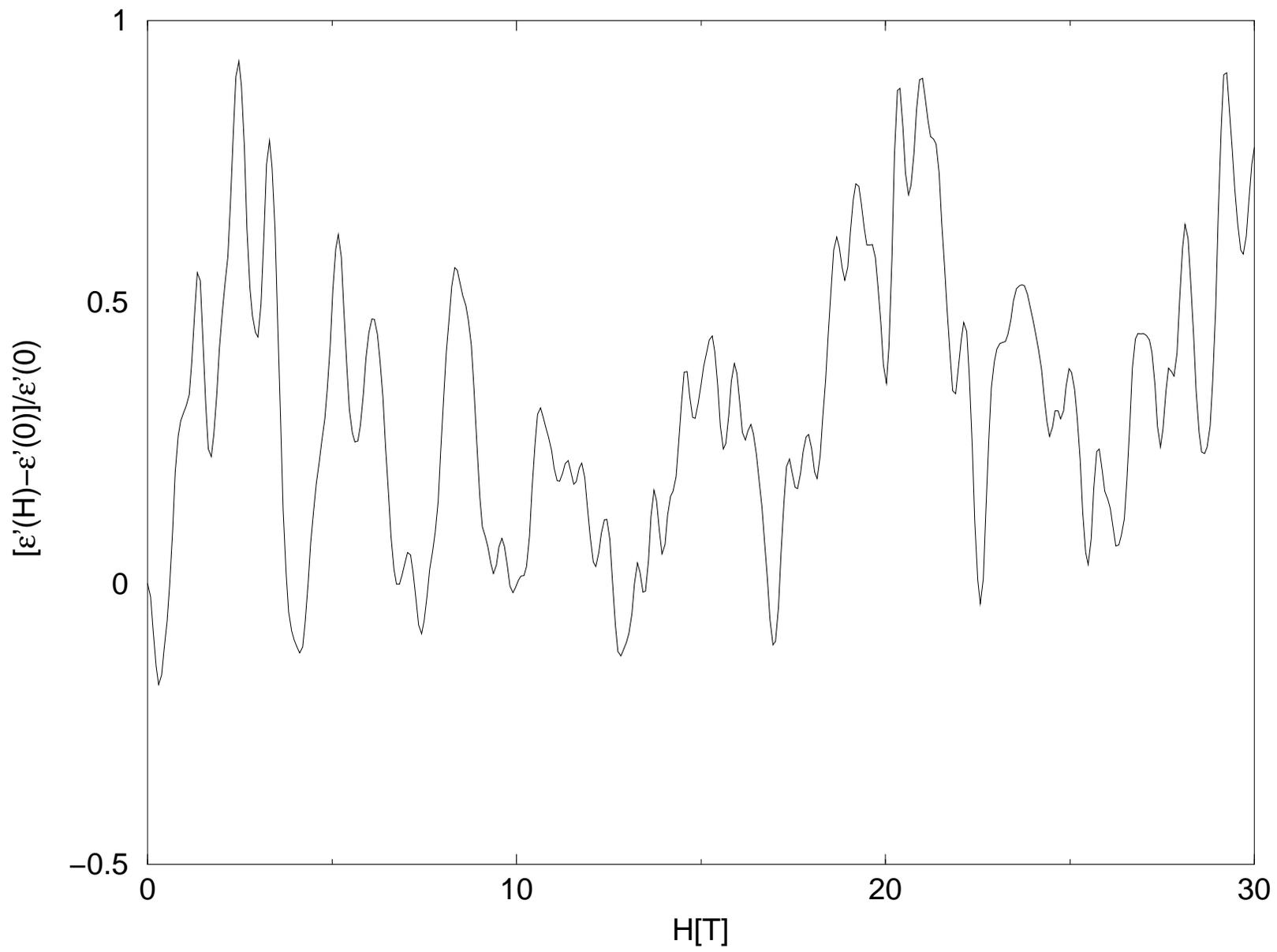
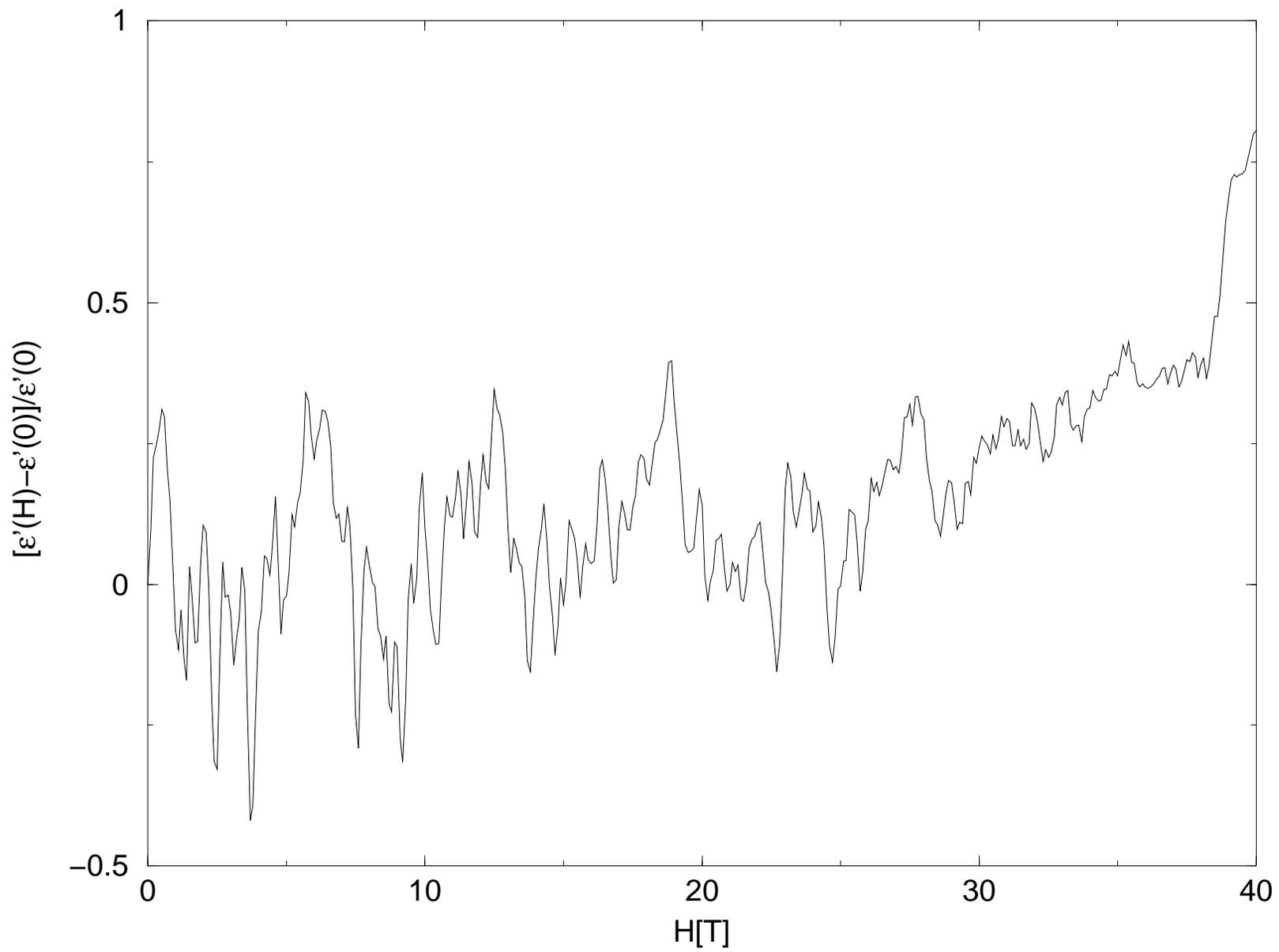
\end{document}